\documentclass{aastex63}

\shorttitle{Chondrule Formation}
\shortauthors{Herbst and Greenwood}
\graphicspath{{./}{figures/}}
\begin{document}

%\title{An Astrophysical Site for Chondrule Formation}
\title{Chondrule Formation During Low-Speed Collisions of Planetesimals:\\ A Hybrid Splash-Flyby Framework}

\correspondingauthor{William Herbst}
\email{wherbst@wesleyan.edu}

\author [0000-0001-7624-3322]{William Herbst}
\affil{Department of Astronomy,
Wesleyan University,
Middletown, CT, 06457, USA}

\author {James P. Greenwood}
\affil{Department of Earth and Environmental Sciences,
Wesleyan University,
Middletown, CT, 06457, USA}

\begin{abstract}

Chondrules probably formed during a small window of time $\sim$1-4 Ma after CAIs, when most solid matter in the asteroid belt was already in the form of km-sized planetesimals. They are unlikely, therefore, to be ``building blocks" of planets or abundant on asteroids, but more likely to be a product of energetic events common in the asteroid belt at that epoch. Laboratory experiments indicate that they could have formed when solids of primitive composition were heated to temperatures of $\sim$1600 K and then cooled for minutes to hours. A plausible heat source for this is magma, which is likely to have been abundant in the asteroid belt at that time, and only that time, due to the trapping of $^{26}$Al decay energy in planetesimal interiors. Here we propose that chondrules formed during low-speed ($\lesssim1$ km s$^{-1}$) collisions between large planetesimals when heat from their interiors was released into a stream of primitive debris from their surfaces. Heating would have been essentially instantaneous and cooling would have been on the dynamical time scale, 1/$\sqrt(G \rho) \sim 30$ minutes, where $\rho$ is the mean density of a planetesimal. Many of the heated fragments would have remained gravitationally bound to the merged object and could have suffered additional heating events as they orbited and ultimately accreted to its surface. This is a hybrid of the splash and flyby models: we propose that it was the energy released from a body's molten interior, not its mass, that was responsible for chondrule formation by heating primitive debris that emerged from the collision. 

\end{abstract}

%% Keywords should appear after the \end{abstract} command. 
%% See the online documentation for the full list of available subject
%% keywords and the rules for their use.


%% From the front matter, we move on to the body of the paper.
%% Sections are demarcated by \section and \subsection, respectively.
%% Observe the use of the LaTeX \label
%% command after the \subsection to give a symbolic KEY to the
%% subsection for cross-referencing in a \ref command.
%% You can use LaTeX's \ref and \label commands to keep track of
%% cross-references to sections, equations, tables, and figures.
%% That way, if you change the order of any elements, LaTeX will
%% automatically renumber them.
%%
%% We recommend that authors also use the natbib \citep
%% and \citet commands to identify citations.  The citations are
%% tied to the reference list via symbolic KEYs. The KEY corresponds
%% to the KEY in the \bibitem in the reference list below. 

\section{Introduction} 

Chondrules are mm-scale igneous spherules that occupy 20-80\% of the volume of nearly all chondrites, the most common type of meteorite. Their existence remains famously unexplained despite a major scientific effort spanning decades \citep{R18}. Unless and until we achieve a better understanding of how chondrules and chondrites formed, it will be hard to be confident about the initial stages of planet formation in the Solar System or how to decode the wealth of information about it available from meteorite studies. 

For decades, models of chondrule formation have been categorized as either nebular or planetary \citep{W63, C05}. Nebular models assume that mm-scale pre-chondrule dust aggregates embedded in a protoplanetary disk of originally solar composition were heated by shocks or some other process and chondrules formed. The compositions and textures of chondrules strongly support this view, which has been the dominant one \citep{HR90,J18}. However, it is now well-established that most chondrules formed in a relatively high pressure oxidizing atmosphere that is difficult to reconcile with a H-rich solar disk \citep{G08, A08, G12, F12, F16, EAL18}. It is also now clear that chondrules formed primarily or exclusively during a narrow interval of time that did not begin until $\sim$1 Ma after CAIs and was over by $\sim$3 Ma later \citep{P19,P23}. The precursors of chondrules in the nebular model are pebble-sized and smaller grains, which aggregate into larger objects on such a short timescale, $\ll0.1$ Ma \citep{W00}, that it is hard to see how they could still be dispersed in a nebular gas in any abundance by the time most chondrules formed.   

Planetary models, by contrast, see chondrules as a byproduct of the collisions of planetesimals, with the required dense, oxidizing atmosphere being created during the collision. In one version,  which requires high-speed collisions ($\gtrsim$2.5 km s$^{-1}$) and is known as the ``jetting" model, there is sufficient kinetic energy in the collision to heat some of the primitive surface material on the colliders to chondrule-forming temperatures, and it has been verified by numerical experiment that this would be expected \citep{JC18}. It has not been demonstrated, however, that objects with chondrule-like textures and other properties could emerge from such jets, and there is no explanation in such a model for why chondrule formation should be limited to $\sim$1-4 Ma.  

Another type of planetary model, known as the ``splash" or ``dirty impact plume" model invokes low-speed ($\lesssim$1 km s$^{-1}$) collisions to release already-molten material from the interiors of large planetesimals \citep{A11, SS12, SS18}. It can be shown that, during the chondrule formation epoch, planetesimals larger than $\sim$50 km in radius would commonly have been substantially molten beneath an insulating outer layer due to the decay of $^{26}$Al, making the model attractive.  Whereas collisions occur at all epochs, there is only a narrow window of time when large planetesimals with molten interiors were common in the asteroid belt, and that time is more or less coincident with the chondrule formation epoch. A fundamental issue with the splash model is whether mm-scale solid spherules will actually emerge from expanding sheets of magma released during such collisions. One has only a sketch of how this might have occured \citep{A11}, with no detailed modeling or experiment to support the idea. A counter-argument to the claim has been made by \citet{L16,L18} who point out that chondrules have relatively iron metal-rich compositions characteristic of primitive material. If their progenitors were molten material inside planetesimals one would expect that at least some of them would have largely been drained of their iron. Such ``basaltic" chondrules are rare in nature and the splash model needs to account for that fact.

A related model that also relies on the existence of hot material near the surfaces of large planetesimals to account for chondrules is the {\it flyby} model, proposed by \citet{HG16, HG19}.  Instead of positing chondrule formation directly from the magma splash, these authors rely on the radiant {\it energy} from magma, released to space by either volcanism or collision, to melt primitive solid material orbiting closely past it during a ``flyby". The pre-chondrule material in this case is primitive dust or m-scale porous boulders. A challenge for this model is to produce enough chondrules, given that it requires the presence of surface lava on a planetesimal at just the time that primitive material is ``flying by". The model is tenable only if a large amount of accretion in the 1-4 Ma time frame is through a circumplanetesimal disk. In that case, the small chance of any individual dust grain or boulder being irradiated by magma from a volcanic eruption or collisional rupture of the carapace during a single orbit, may be multiplied by the millions of orbits it may make before accreting to the surface of the planetesimal. If most accretion is not through a disk then few chondrules could be formed in this way. 

The lack of progress to date in accounting for chondrules is not due to a lack of observational constraints -- there are now many \citep{CJ16, R18}. \citet{P19} enumerate six: peak temperatures, cooling rates, elevated alkali vapor pressures, chondrule reworking at high temperatures, matrix-chondrule complementarity, and a chronology that places chondrule formation in the context of planet formation. Other constraints include the magnetic properties of chondrules and chondrites \citep{E11, F14, F18, S17}, the phenomenon of size sorting \citep{F15}, the existence of compound chondrules \citep{GK81, W95, AK05, J21}, cluster chondrites and ``hot accretion" \citep{M12,B19}, and the overall abundance of chondrules in the near-Earth Solar System today (discussed below). 

In this paper we propose a framework for chondrule formation that is a hybrid of the splash and flyby models. We point out that the dirty impact plumes of the splash model are an attractive astrophysical site for forming chondrules by the flyby mechanism -- radiative heating of primitive material orbiting close to the surface of a large planetesimal -- although in this case the proto-chondrules might be flying ``with" the heat source (magma) rather than ``by" it. Our proposal differs from the splash model in that chondrules are not presumed to form directly from the magma itself, but from primitive debris heated by the magma. It differs from the flyby model in that a single event -- a low-speed collision -- is responsible for both releasing the $^{26}$Al decay energy stored in  planetesimal interiors and generating the mm- to m-sized primitive objects on ballistic orbits to be irradiated. In Section 2 we describe the idea and its justification in more detail. In Section 3 we discuss how the hybrid framework not only addresses the primary inadequacies of the splash and flyby models, but offers a promising avenue for meeting the full range of constraints on chondrule formation. Section 4 is a brief summary.  

\section{The Case for a Splash-Flyby Hybrid Model}

The fundamental issue in explaining chondrules is to identify a heat source strong enough to melt solids that could also be modulated on a timescale of minutes. The seeming plausibility of any particular scenario depends on one's views of the chondrule formation environment and the efficiency with which they needed to be produced. We begin, therefore, with a review of the evidence suggesting that chondrules formed in the asteroid belt between 1-4 Ma after CAIs, at a time when most of the original solids had already formed into planetesimals large enough to have molten interiors. In that case, chondrules were not essential to planet formation, were never abundant on asteroids, do not require an efficient production mechanism, and there is an obvious heat source -- magma encased in planetesimal interiors. It has the right temperature and is present in abundance at the right time and place to serve as the heat source. In the remainder of the Section, we sketch a proposal for how this might have occurred.    

\subsection{Where and When did Chondrules Form?}

 We assume that chondrules formed in the asteroid belt, close to the locations where their host chondrites are currently stored on surviving primitive asteroids \citep{P21}. The more volatile-rich carbonaceous chondrites (CCs) presumably formed at the outer edge of the belt and have isotopic characteristics distinctly different from the ordinary chondrites (OCs) that formed closer in and, perhaps, slightly earlier \citep{F22}. The distinctive isotopic characteristics of the CCs and OCs is sometimes attributed to an early formation of Jupiter, which would dynamically stir the planetesimals generating high speed collisions, but may instead reflect other aspects of the evolution of the protosolar disk, such as the location of the water ice line and/or the development of pressure bumps that constrain the motion of pebbles \citep{L21, I22}. \citet{D21} simulate the formation of Jupiter and find that it is too slow to be a significant dynamical influence on the asteroid belt until $\sim$3.4-4.2 Ma. 

The Al-Mg dating technique has been applied to over 150 chondrules from a wide range of meteorites \citep{V09, KU12, NKL18, P19, S22}. \citet{P19} interpret the results as indicating a rather abrupt turn-on to the main chondrule formation epoch at about t = 1.8 Ma, where t = 0 is set by the $^{26}$Al abundance at the time of CAI formation, followed by some possible peaks at 2.0 and 2.3 Ma and an end to the process by $\sim$3 Ma. Allowing for some outliers one might characterize the epoch broadly as between t = 1.5 - 4 Ma with a peak near 2 Ma. A more recent analysis by \citet{P23} suggests that the gap between CAI formation and the start of chondrule formation may be as small as 0.7 Ma, but it remains clear in their data that CAIs formed in a separate heating event or events that significantly pre-dated the main epoch of chondrule formation.  

One key aspect of this description that is vigorously debated is whether any surviving chondrules at all formed before t$\sim$0.7 Ma. There are some chondrules whose Pb-Pb ages appear to overlap with those of CAIs \citep{CB12, B17, CB18}. One interpretation of this is that the chondrule formation epoch actually began at t = 0, coincident with CAI formation and, since the 22 chondrule ages determined by this technique are roughly evenly distributed over the 0 - 4 Ma epoch, that chondrule formation was steady over this time. In the model proposed here, no chondrule should be as old as CAIs, assuming they date the beginning of planetesimal growth,  since it does take at least a few hundred thousand years for large planetesimals to form and generate molten interiors \citep{HS06}, but chondrule formation could have begun earlier than 0.7-1.8 Ma. Given that the canonical $^{26}$Al-abundance is about four times larger than is necessary to fully melt planetesimals larger than 5 km or so \citep{SS18}, chondrules that formed earlier than 0.7-1.5 Ma (about 1-2 half-lives of $^{26}$Al) may simply have not survived further processing. 

Ways of reconciling the discordant Al-Mg and Pb-Pb results are discussed in several papers \citep{NKL18, CB18, Pi19, P19, D22, P23} and need not be elaborated upon here. We simply note that there is no plausible mechanism proposed in these papers for producing a heterogeneity of $^{26}$Al at the level required and that accounting for all possible error sources in the Pb-Pb ages, including possible systematic effects from environmental contamination, is challenging. If the error bars on the individual Pb-Pb ages are underestimated, then there may actually be no discordance. The average time for chondrule formation by both techniques is about 2 Ma after CAIs, or 1 Ma if the analysis of \citet{P23} is correct.  

\subsection{The Implausibility of Nebular Models}

Any successful model of chondrule formation must be consistent with what is known about the early evolution of gas and solids in the Solar System. It is widely accepted that dust grains aggregate to m-scale objects and settle towards the disk plane of the solar nebula on a very short timescale, $< 0.1$ Ma \citep{W00, CY10}. Therefore, the progenitors of chondrules in nebular models -- mm-scale dust -- are likely to be rare by 1 Ma. Even the nebular gas in the 1.5-4 AU zone may be seriously depleted by 1 Ma, since that is roughly the half-life of disks around solar mass stars inferred from observations of young stars in the solar neighborhood \citep{M09,V17}. And, despite decades of effort, no mechanism for heating nebular gas to the requisite temperatures for the requisite times has yet gained widespread acceptance. Global nebular models would be required if a large percentage of the original solids in the disk needed to be turned into chondrules but, as we argue further below, that is not the case. 

\subsubsection{Simulations of the Evolution of Solids and the Chemistry of the Ambient Gas}

It is increasingly clear that planetesimals large enough to contain their $^{26}$Al decay energy emerged quickly in the proto-solar nebula with the result that objects with molten interiors were common. Observational evidence for this comes from the inferred ages of iron meteorites \citep{Scott20}. Additional evidence comes from simulations of planetesimal growth in the asteroid belt during the first 2 Ma by \citet{W19}. He assumes that bodies grow by accretion alone from primary planetesimals in the 50 m to 5 km range and adopts an initial mass of solids between 1.5 and 4 AU of $\leq4.9$ M$_\oplus$, consistent with a smooth surface density variation in the solar nebula. He further assumes that Jupiter has not yet fully formed and that any dynamical effects from it can be neglected. 

One purpose of these simulations was to test whether the large, differentiated planetesimals (LDPs) needed to account for iron meteorites could form quickly enough to satisfy timing constraints imposed by their radioactive ages. This turns out not to be a problem; for an initial planetesimal radius of 50 m, tens of thousands of objects exceeding 50 km form quickly and then continue to accrete at a slower pace. Somewhat counter-intuitively, he also finds that the smaller the initial (primary) planetesimals, the faster the evolution proceeds. For the 50 m simulation, the fraction of mass in  planetesimals of different size, based on that simulation, is shown in Fig. 1 at t = 1 Ma and 2 Ma. We note that the starting points, t = 0, for the simulation and for the radioactive age scale are defined differently but assume that the actual time difference is negligible for the purposes of the discussion in this paper. The size distribution is truncated at 500 km because objects of this size and larger are assumed to be ejected from the Solar System once Jupiter forms, at perhaps 3-4 Ma. Again, nebular models require mm-scale objects to be embedded in the disk gas and observation, theory and simulations suggest that would not be the case in the asteroid belt at 1 Ma.    

\begin{figure}[ht!]
\plotone{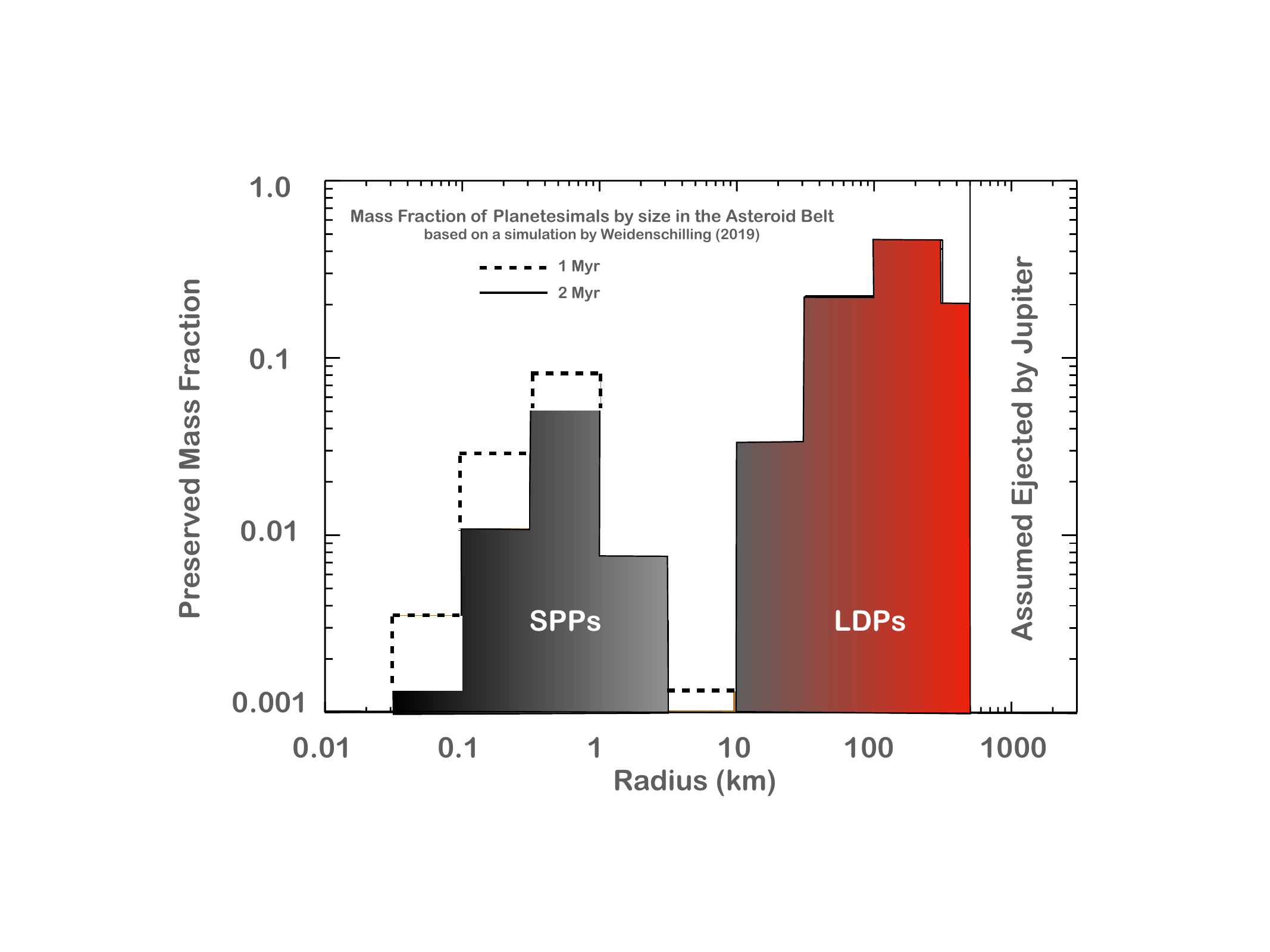}
\caption{The mass fraction of planetesimals as a function of size in the asteroid belt at 1 Ma (dashed lines, except in case of the 1-3.1 radius bin, where the dashed and solid lines overlap) and 2 Ma (solid lines) based on a simulation by \citet{W19} (see his Fig. 20). The distribution is bimodal at both times. Between 1 and 2 Ma one sees a decline in the mass fraction of primitive matter from about 10\% to 8\% as SPPs (small, primitive planetesimals) accrete to LDPs (large differentiated planetesimals). The accompanying rise in the mass fraction of LDP's is not noticeable on the figure due to the logarithmic scale. The larger LDPs are nearly fully molten (red) while smaller ones may retain substantial primitive material in surface layers (indicated by red to black shading). For this simulation it was assumed by \citet{W19} that Jupiter formed after a few Ma and gravitationally ejected a few very massive (r $\ge$500 km) objects, along with numerous smaller objects that carried much less mass. The net result of Jupiter's action would be to enhance the ratio of primitive to differentiated mass that was present at 2 Ma and is ``preserved" in today's Solar System from 0.25\% to 8\%.}
\end{figure}

There is another line of evidence that favors a planetary, as opposed to nebular, origin for chondrules. While calculations of oxygen fugacity (fO$_2$) suggest CAIs solidified in a reducing gas, as expected for a nebular disk \citep{G08}, FeO-poor chondrules formed in an O-rich gas, $\sim$5 orders of magnitude higher in fO$_2$ \citep{G12}. This is a challenging fact to explain for any nebular model of chondrule formation, including those invoking shock waves \citep{F12}. \citet{F16} show that CI-dust enrichment by about 4 orders of magnitude and water-enrichment are required for FeO-rich chondrules and point to impacts on water-rich parent bodies as a potential source of the gaseous conditions indicated. \citet{A08} used measurements of Na abundance and zonation in chondrules to argue that they must have formed in a gaseous environment with a high partial pressure of Na, so high in fact that if the system were of solar composition, these regions would have been self-gravitating. Similar arguments can be made with respect to Fe and all moderately volatile elements present in chondrules \citep{EAL18}. If chondrules formed in a typical proto-planetary disk environment their more volatile elements would be expected to have evaporated, resulting in elemental depletion and isotopic fractionation effects that are not observed. The conclusion of these authors is that chondrules formed in regions of space where the density of solids was much higher than current nebular models can explain.

\subsubsection{Can a Planetary Setting Produce Enough Chondrules to Explain their Prevalence in Chondrites?}

An early objection to all planetary chondrule formation models was that they cannot explain the seeming ubiquity of chondrules within primitive material. If chondrules were as abundant in the early asteroid belt as they are in chondrites, where they make up $\sim$50\% of the mass, then they would need to be made with high efficiency. However, it is increasingly clear that the Earth's atmosphere filters out a substantial amount of poorly-lithified (and arguably chondrule-poor) primitive material that is present in space, so that the actual mass fraction of primitive material in the form of chondrules near the Earth today is probably more like 0.01-0.1\% than 50\%. This point was first made by \citet{Se98}, who further noted that the transport processes needed to bring material from the asteroid belt to 1 AU may also bias the near-Earth sample in the direction of chondrule-rich material. 

Recent results, including improved data on near-Earth asteroids and sample returns, support the view that chondrules are probably much rarer in space than they are among meteorites in the world's collections. Based on the known meteorite flux of 2900 - 7300 kg a$^{-1}$ \citep{B96}, and the fractional abundance of chondrites among meteorites, 85-90\% \citep{C17}, the flux of chondrules at the Earth's surface is 1.3 - 3.3 $\times$ 10$^3$ kg a$^{-1}$. This may be compared to the total mass flux at the top of the atmosphere, 1 - 6 $\times$ 10$^7$ kg a$^{-1}$, derived from meteor rates, impacts on space hardware and other studies \citep{D17}. Chondrules, therefore, are only a minor component of the material impacting the Earth at the top of the atmosphere, $\sim$0.01\% of the mass. Most of the impacting mass arrives in particles much smaller than chondrules. It can be collected from space or the upper atmosphere as IDPs and from the surface as micrometeorites and does not resemble ground-down chondrules, in general being much more porous \citep{KS19}. Whether it is primarily from comets or from the asteroid belt is unknown.

Another avenue for directly assessing the fractional abundance of chondrules within chondritic material has recently opened, namely sample returns from the surfaces of small asteroids. Regrettably, the Hayabusa mission did not return enough material from the surface of Itokawa, an asteroid chemically similar to LL-class OCs, to determine its chondrule abundance or even whether any chondrules at all are present on its surface \citep{A06}. Hayabusa2 was much more successful, however, returning several grams of material from Ryugu. The first analyses show that it is definitely composed of chondritic material with zero chondrules present \citep{Y21}. \citet{N22} describe some very small particles in the sample as related to chondrules but the relationship is uncertain. The rarity or complete absence of chondrules in the Ryugu sample was predicted by \citet{HG21} on the basis of its low bulk density, 1.19 g cm$^{-3}$, and its low macroporosity, $0.14 \pm 0.04$, calculated from linear-mixture packing theory \citep{YS91}. They showed that, if the asteroid was reasonably homogeneous, its surface rocks must be much less dense than any known meteorite, even the rare chondrule-free CI class. Such low-density, micro-porous material is consistent with the thermal properties of Ryugu \citep{G19}, and argues against the presence of many chondrules in the material, since their densities are in the range 3.0 - 3.4 g cm$^{-3}$ \citep{F15}. \citet{HG19} also correctly predicted, to within the margin of error, the mean density of the particles returned by Hayabusa2 \citep{Y21}, and it is significantly less than that of any known meteorite. 

Could Ryugu be an unusual primitive asteroid with regard to its chondrule content, or is it likely to be representative? A definitive answer will await further results, but we note that a sample from Bennu obtained by the OSIRIS-REx mission has recently arrived on Earth and that the bulk density of this asteroid is also 1.19 g cm$^{-3}$ \citep{L17}. The general appearance of its surface is also quite similar to that of Ryugu and even without comparing their boulder-size distribution in detail, linear-mixture theory suggests that the macroporosity and, therefore, the mean boulder densities of the two asteroids will be much the same. That is because the macroporosity of a granular aggregate depends primarily on the dispersion of grain sizes of its component boulders, not on their actual size or even shapes, as long as they are not extreme in some way. As \citet{YS91} discuss, single-sized grain piles typically have a porosity of around 40\% but two factors cause the porosity to decline if there is a range of grain sizes. Larger than average particles displace void space, reducing porosity, a phenomenon known as occupation, whereas smaller than average particles fill in void spaces, a phenomenon known as sifting. Both effects act to lower the porosity of the pile. The dependence of porosity on the range of grain sizes for a lognormal distribution, which represents Ryugu very well \citep{HG21}, is shown in Fig. 2. Given the appearance of the surface of Bennu, with its wide range of boulder sizes, linear-mixture theory indicates a macroporosity of about 0.15. We, therefore, predict the OSIRIS-REx sample will have particles similar in density to Ryugu and contain few, if any, chondrules.  

Is it possible that Ryugu and Bennu are not representative of primitive asteroids in general because they are spectrally similar to CC's rather than OC's and CC's are only 5\% of the meteorites in our museums? The best studied S-type asteroid is Itokawa, which was visited by the original Hayabusa mission. Unfortunately, its sample return mechanism did not function properly so we do not know if it would have brought back chondrules or not. However, given the asteroid's low bulk density of $1.9 \pm 0.1$ g cm$^{-3}$ \citep{A06}, and assuming a macroporosity of 0.15 (Fig. 2), the predicted average density of its boulders is only 2.2 g cm$^{-3}$, much less than the density of its analog meteorite class (LL; 3.2 g cm$^{-3}$) again leaving little room for abundant chondrules. Of course, if chondrules did not begin forming until t = 1 Ma, when the planetesimals that evolved into asteroids were already large, we should not expect to find chondrules in any abundance on any asteroid. To summarize, a variety of arguments suggests that chondrules are not nearly as abundant on asteroids as they are in chondrites and, therefore, a chondrule formation mechanism need not be very efficient at making them. Based on current knowledge, models should be judged on their ability to process $\sim$0.01 - 0.1\% of relevant material into chondrules, meaning that global (nebular) models are not required. 

\begin{figure}[ht!]
\plotone{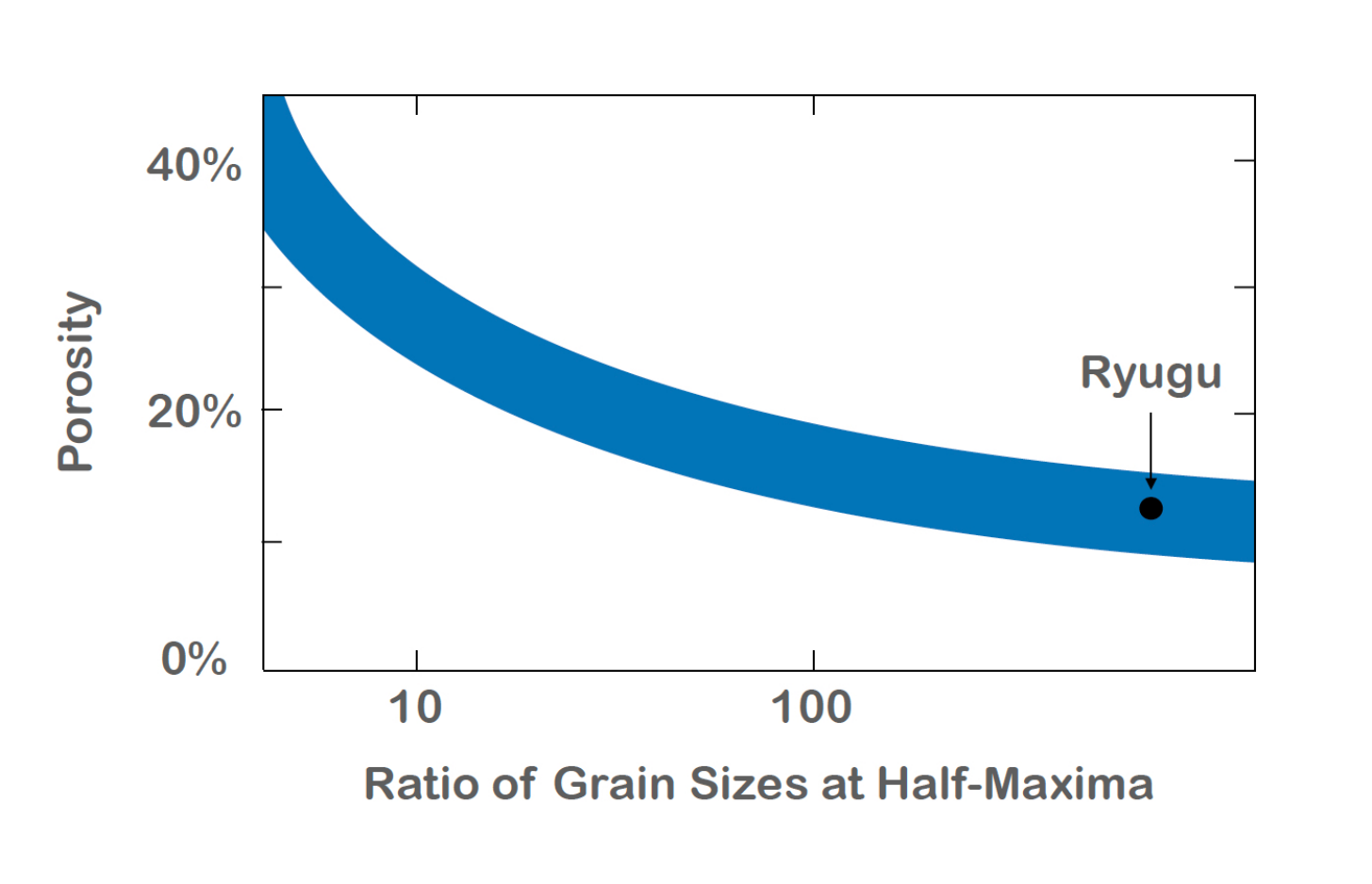}
\caption{A schematic figure showing how increasing the size range of grains in a granular aggregate decreases the porosity of the pile. The curve displayed is for a lognormal distribution \citep{YS91} but the principle applies to any physical system in which occupation and sifting operate. The location of Ryugu, based on the surface boulder counts of \citet{M19} and the analysis of \citet{HG21} is shown. It has FWHM = exp ($2 \sqrt{2 {\rm ln} 2}\ \sigma$) = 285, which means that the grain sizes of particles at half-maximum on the high side of its frequency distribution are 285 times larger than those at half-maximum on its low side. For Ryugu this corresponds to $\sim$20 m boulders on the high side and $\sim$7 cm cobbles on the low side. The width of the band is set by the fact that, excluding extreme shapes or non-gravitational forces, piles of single-size grains typically have a porosity of $40\% \pm 5\%$.}
\end{figure}

\subsection{Jetting versus Magma as a Heat Source in Planetary Models}

An alternative heat source to magma in planetary models is the kinetic energy of the collision. Material could be raised to the melting point of solids by high speed  collisions, but it requires impact speeds of 2.5 km s$^{-1}$ or larger \citep{JC18}. In the \citet{W19} simulations, such high speeds do not occur because objects large enough to generate them have not yet formed. Impact speeds in his model would be at roughly the escape velocity of the larger planetesimal, which is $$ v_{esc} = \sqrt{ {2 G M} \over R} = \sqrt{ {8 \pi G \rho} \over {3}} R = {{1.3}\  \big( {{R} \over {1\ \rm{km}} } \big)}\ \rm{m}\  \rm{s}^{-1} $$ where {\it M}, {\it R} and $\rho$ are the mass, radius and mean density of the larger planetesimal and we have adopted $\rho = 3000$ kg m$^{-3}$. As Fig. 1 shows, most collisions will be between objects with $R \le 300$ km, so typical collision speeds will be under 0.4 km s$^{-1}$, insufficient to form chondrules by jetting. If, contrary to an assumption on which the \citet{W19} simulations are based, Jupiter had formed by the beginning of the chondrule formation epoch then a jetting model would be viable from 1 Ma on. However, a question would remain -- why does chondrule formation end, at about 4 Ma? High speed collisions between planetesimal-sized objects would presumably continue beyond that time and, if such collisions can generate chondrules between 1 and 4 Ma, why would they not do so at later times? And, of course, it has not been shown experimentally or by simulation that objects with the distinctive properties of chondrules could emerge from any high speed collision.

\citet{HG16, HG19} proposed near-IR radiation from magma as the heat source for chondrule formation, noting that it has the right temperature and is arguably quite abundant in the asteroid belt during the chondrule formation epoch. \citet{A11} and \citet{SS12} proposed that chondrules actually form directly from such hot magma. Thermal evolution calculations indicate that objects with  $R \ge 50 $ km will be largely molten within $\le$0.5 Ma of their formation, assuming a canonical value for the initial $^{26}$Al abundance \citep{HS06, SS12, SS18}. According to the \citet{W19} simulations more than 90\% of the original mass of the asteroid belt was already encased in planetesimals large enough to grow substantial molten interiors by t = 1 Ma. The ages of iron meteorites provide additional support for the prediction that molten interiors were a feature of planetesimals at that time \citep{Scott20}, as noted above. By about 1.5 Ma, the $^{26}$Al level would have dropped to the point where cooling would dominate heating, and by the end of the chondrule formation epoch at 4 Ma, one can expect that the amount of magma in the asteroid belt would have diminished significantly and the insulating carapaces grown thick enough that remaining interior heat would rarely be exposed to space. Fig. 3 is a schematic representation of the situation.

\begin{figure}[ht!]
\plotone{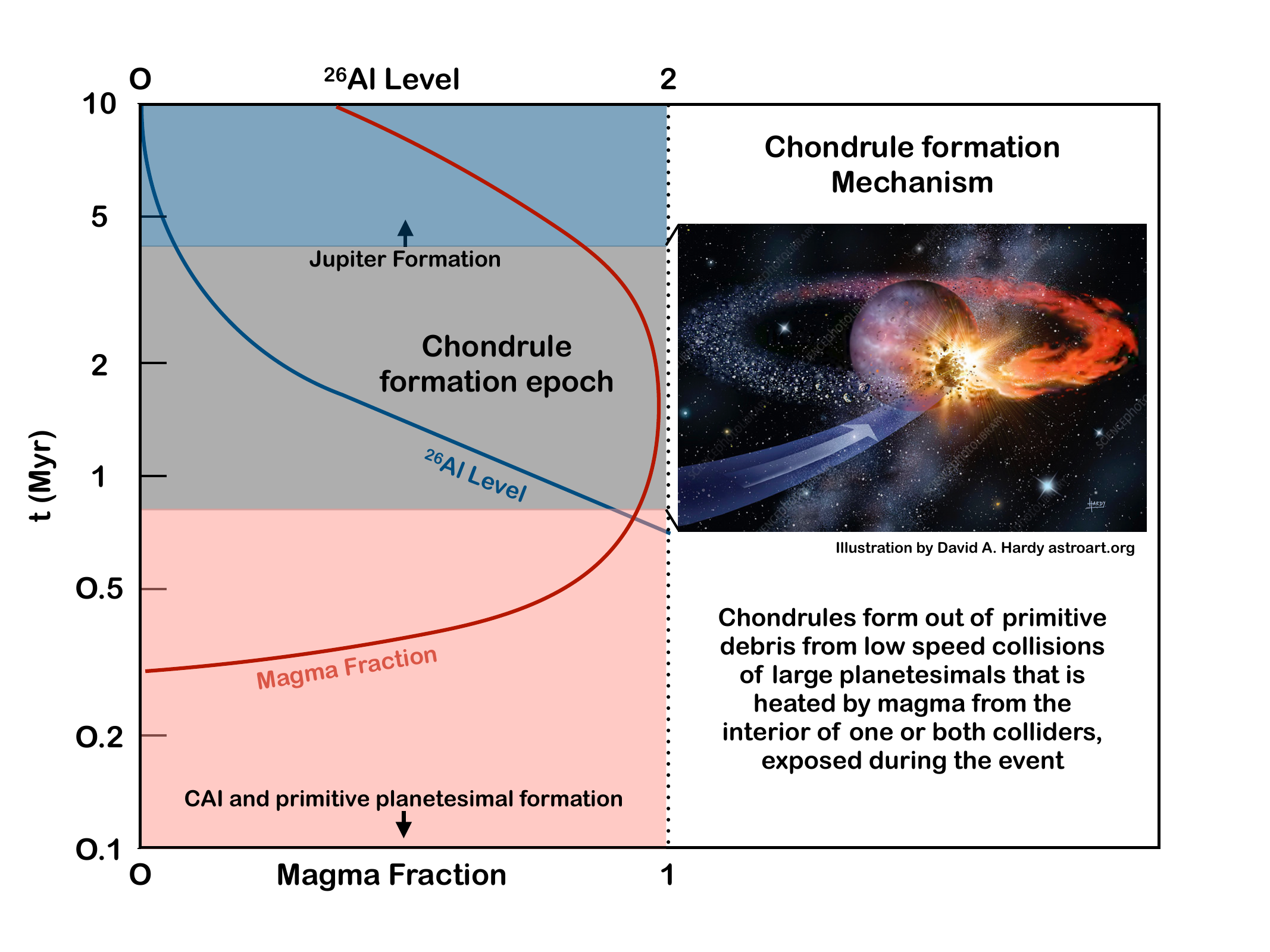}
\caption{A schematic chronology of the early asteroid belt based on the evidence discussed in the text. CAI formation defines t = 0 and primary planetesimals are assumed to form quickly. Within a few hundred thousand years most of the mass is in the form of large planetesimals, which become dominated by magma interiors, as shown by the red line labeled magma fraction. The energy source is $^{26}$Al, whose level declines with a half-life of 0.717 Ma from an assumed initial value of 4 at t = 0, where a level of 1 corresponds to the amount of energy (1.6 kJ/g) needed to just reach the liquidus of a perfectly insulated cold primitive starting material. Chondrules formed between $\sim$1-4 Ma when exposed to hot magma released during a collision, as illustrated by the drawing in the right hand panel by David A. Harvey (astroart.com) originally depicting the formation of a ring around the Earth. Chondrule formation ends when the LDPs have cooled sufficiently that magma does not reach their surfaces even during collisions. Jupiter may form at around 3-4 Ma and begin clearing the asteroid belt of many of its larger objects.}
\end{figure}

\subsection{The Hybrid Model}

There are two planetary models that rely on the trapped energy from the decay of $^{26}$Al as the heat source -- the splash and flyby models, and each faces challenges. As noted in the Introduction, there is little evidence to support the idea that chondrules would form in a splash of magma and the rarity of basaltic chondrules is a problem for the model. Relict grains and dusty rims are observed features of some chondrules that are also challenging for the splash model (or jetting model) to explain. Even if mm-scale drops of molten material do emerge from ejecta sheets, they must absorb primitive material from the ``dirty plume" in the proper amounts to account for these observed features of chondrules. It seems that the splash model involves two so-far unverified assumptions: the creation of mm-sized molten drops from expanding sheets of magma and the absorption by those drops of primitive material during the splash. 

By contrast, in the flyby model chondrules formed conventionally during a brief, intense heating of primitive material and magma was involved only as the heat source. But the required radiation can only be liberated to space briefly and episodically by rupturing the surface of an LDP and the pre-chondrule material (primitive dust or porous blocks) must be within kilometers of it to receive a sufficient dose of energy to melt. \citet{HG19} envisioned a volcanic magma extrusion on the surface of an LDP, modeled as a circular hot spot, irradiating orbiting material within a circumplanetesimal accretion disk. They calculated the heating/cooling curves expected in such a case and demonstrated via laboratory experiments that objects with realistic chondrule textures could be synthesized from mineral precursors, building on the classic work of \citet{HR90} and complementing the more recent studies reviewed by \citet{J18}. 

While there is little doubt that chondrule-like objects could have been made in this manner, it would have been exceedingly inefficient unless there were a mechanism to break primitive material into mm- to m-size objects and accrete them through a circumplanetesimal ring, since any surface extrusions of lava would be rare and short-lived. One way to do this would be by tidal capture and disruption of small, primitive planetesimals. Quantitative analysis of this mechanism indicates, however, that it is not likely to be important during the kind of planetesimal growth phase simulated, for example, by \citet{W19}; it is not part of his simulations. The other possibility is that a single event, an oblique collision of two planetesimals, both created the swarm of m-scale or smaller debris to be irradiated and released the requisite energy stored in the interior of one or both colliders. It is this hybrid of the splash and flyby models, that we propose here as worthy of more computational attention, and discuss below in qualitative form.   

The idea is based on the oblique collisions described by \citet{A11} and \citet{A23} that are a basis for the splash model. As the authors point out, in the 1-4 Ma time frame there will be many collisions between large planetesimals, and most will be sufficiently oblique that a debris field will emerge downstream from the event. If one or both of the colliders has a large molten interior below a primitive surface the debris field may contain both incandescent magma and primitive solids. High spatial resolution simulations are not available to confirm the size distribution of the primitive debris but it is reasonable to expect it will include everything from dust to boulders. These objects will be on ballistic orbits, most of which will be closed. A visual representation of such an event, originally inspired by a model of the formation of the Moon, is included in Fig. 3. 

Based on the simulation by \citet{A11} one might expect that shortly after the collision, primitive material with a range of sizes will find itself immersed in an expanding,  $\sim$2000 K black body radiation field whose source is the magma spray within the debris field. In a matter of minutes one can expect significant cooling, as the fan of material expands, becomes optically thin and the magma cools and solidifies, perhaps as small droplets \citep{A11}. Although we are not aware of any simulations that are sufficiently detailed to predict heating/cooling curves in such a scenario that could be tested in the laboratory, it seems likely to us that they would be quite similar to what has been modeled successfully by \citet{HG19}. In particular, the initial fan-shaped spray of hot magma from the interior(s) should have a peak temperature exceeding 1600 K, which is what is required to form chondrules. It should then cool on a timescale of minutes as it expands  \citep{A11}. Much of the debris will presumably remain bound to the merged planetesimal and orbit on the dynamical timescale of 1/$\sqrt(G \rho) \sim 30$ minutes. If the magma interior of the merged planetesimal remains exposed, there could be multiple episodes of heating as this material orbits past the impact site every 30-60 minutes or so. Eventually, much of the processed debris, would be expected to accrete to the surface of the merged planetesimal, as chondrules, matrix grains condensed from the cooling vapor and perhaps even already-lithified chondrites. Without higher resolution simulations covering a wider swath of parameter space it will not be possible to test this idea quantitatively but, qualitatively, it seems promising to us as a framework for understanding many of the observed constraints on chondrule and chondrite formation, as we now discuss.    

\section{Discussion: Testing a Splash-Flyby Hybrid Framework}

There is a substantial number of strong constraints on the formation of chondrules and chondrites, a fact that has led some to describe the problem as ``over-constrained" and others to assert that no single model of chondrule formation is likely to be able to account for all of them. In this section we briefly describe how the hybrid splash-flyby framework proposed here may be a promising one for addressing these constraints. Unless and until numerical simulations of the proposed events are carried out, it is only possible to make qualitative arguments. 

\subsection{Density of Solids in Chondrule Formation Zones}

It has been recognized for decades that some chondrules must have interacted with at least one other chondrule while both were still hot and plastic \citep{GK81, W95, AK05}. Such objects are referred to as compound chondrules and their existence provides an important constraint on chondrule formation since it proves that the number of chondrules forming within a given volume at the same time must have been high enough that soft collisions could have occurred. The work referred to above found that the fraction of chondrules suffering such collisions was of the order of a few percent. Recently, \citet{J21} has argued that the actual fraction of chondrules suffering an interaction with another chondrule during their formation is much higher than previous estimates. Based on his analysis of lobate chondrules in two CO3 meteorites, he argues that the percentage of chondrules that could be called compound, because they interacted with another chondrule while both were still in the process of formation, has been underestimated by an order of magnitude or more. He argues that, in fact, almost all chondrules may be compound. If true, this places a constraint on the density of solids in the chondrule formation zone that is is difficult to account for by any nebular model. The dust to gas ratio would need to be increased by such a large factor that there would be insufficient gas remaining to heat the dust sufficiently by any mechanism for chondrule formation. In the hybrid model prosoed here, gas density is irrelevant and the formation of molten and partially molten droplets that will solidify into chondrules can occur with whatever concentration of them is needed to allow their interaction and formation into the number of lobate chondrules observed. There is no need to bring hot chondrules together from any substantial distance, since the proto-chondrules are already in close proximity to one another.  

\subsection{Magnetic Properties}

Some chondrules retain a record of the strength and direction of the magnetic field at the time their iron-bearing minerals last cooled through their Curie points, typically $\sim$1000 K \citep{F18}. Field strengths recorded are in the range 10 - 100 $\mu$T and field directions may or may not be coherent across the meteorite. The detection of these natural remanent magnetizations shows that there was a stable magnetic field present at the time and location of chondrule formation and that it was weaker than predicted by some models of chondrule formation, such as those relying on electric currents or being close to the Sun for heating \citep{F14}. If the fields detected can be confidently attributed to the solar nebula then inferences about its nature can be drawn \citep{B20}. However, another likely source of magnetic fields in the chondrule formation zone and epoch is an LDP dynamo \citep{E11}. They are commonly invoked to explain meteorite-wide alignments of the field on a parent body after chondrite formation. In the model proposed here, all chondrule formation occurs close to an LDP so the magnetic field source could be an LDP, and no nebular field would be required. The pre-chondrule material may also be packaged in larger units, meter-scale blocks, than single chondrules and the predicted cooling rates near 1000 K are likely to be high, up to 6000 K/h. Both of these factors increase the probability that an individual chondrule could remain sufficiently stationery as it cools through its Curie point to retain a magnetic record of the field. Moreover, the fact that chondrite lithification may occur simultaneously with chondrule formation (see subsection below on chondrite lithification) allows for the possibility of meteorite-wide coherence of magnetic field direction without having to postulate a second episode of heating. If this view is correct, the remanent magnetizations detected in meteorites would be telling us about the dynamos of large, primitive asteroids, not the nebular disk. This is one reason that it remains important to understand chondrule formation even if they are not the building blocks of planets.

\subsection{Complementarity} 

It has been known for decades that a complementary chemical relationship exists between the chondrules and matrix in chondrites \citep{PHE15}. While the bulk composition of chondrites matches expectation based on the Sun closely, their two main components, chondrules and matrix, do not. Volatile elements are depleted in the chondrules and enhanced within the matrix, while more refractory elements show the opposite trend \citep{H18}. Any successful chondrule formation model needs to account for this widely-observed phenomenon. Isotopic complementarity also may exist \citep{B16}, however see \citet{SS22}. 

Complementarity appears to demand the confinement of atoms released by melting chondrules within a gas from which matrix grains later condense. The confinement could be immediate and local, if there is a mechanism to trap them. It could also be on a scale of AUs if the evaporates escape back into the solar disk but are confined within a ring by pressure gradients. If most of the original solids in the solar nebula were processed through chondrule melting, then the evaporated atoms might be sufficiently abundant that later-condensing matrix grains could bear the signature of complementarity. However, if only a tiny percentage of primitive material ever gets processed into  chondrules, as argued above, then their evaporated atoms would be heavily diluted by diffusion into the nebular gas and it is hard to see how a complementarity signal could survive. Assuming chondrules are actually a rare form of primitive solids then one requires some immediate confinement of vapor released during their formation followed by rapid condensation into matrix grains to explain complementarity. The hybrid splash-flyby model described in this paper offers two possibilities for such confinement. There will be a relatively high pressure, oxidizing gas present during the collision, derived from the more volatile solids. It could serve to limit dispersal of atoms released by chondrule formation sufficiently that they could recondense as matrix grains during the expansion and cooling phase of the cloud following the collision. Alternatively, as \citet{HG19} have proposed, the gas confinement could be by the unmelted portions of meter-scale blocks of primitive material inside of which the chondrule formation occurs. In either case, it appears that the demands of complementarity could be met with the model of chondrule formation proposed here.

\subsection{Cluster Chondrites and Hot Accretion}

Cluster chondrites are portions of OCs that are so densely populated with chondrules that their shapes depart from the normal spherules and take complex forms dictated by the chondrules around them. It appears that these parts of the OC formed and solidified while their chondrules were still warm and pliable, allowing them to mold into the spaces available for them. \citet{M12} has described these objects and reports that they are found in all types of OC and that they are quite common. The implication of cluster chondrites is that the lithification of the chondrite into its current structure must have occurred very shortly after the formation of the chondrules, since they needed to still be hot and pliable. Given that cooling times for chondrules must be minutes to hours, this is a very significant constraint on what happened. Recent work by \citet{B19} has confirmed these findings and expanded the evidence for what is often called ``hot accretion". That descriptor is meant to signify that the chondrules must still have been hot when they accreted to the parent body on which the meteorite formed. In the model described here, still-warm chondrules and just-formed matrix grains could accrete to a portion of the surface of the merged planetesimal that was not molten and chondrite lithification could occur there. Another possibility, as discussed by \citet{HG19} is that chondrite lithification could occur while material was still in orbit, through hot isostatic pressing (HIPping). In either case, the formation of multiple chondrules in close proximity to one another and their rapid incorporation into a chondrite does not seem surprising or difficult to account for in terms of the model of chondrule formation proposed here. 

\subsection{CAI and Pre-solar Grain Storage}

Most chondrites contain CAIs and other highly refractory inclusions, including pre-solar grains, which formed 0.7 Ma or more before the chondrules \citep{M05, M03, W22, P23}. CAIs are believed to be condensates of a brief, hot phase of the solar nebula, although the cause and spatial extent of this hot phase are uncertain. Assuming they were present in the asteroid belt at t = 0 they would have accumulated like everything else into km-sized planetesimals and could have survived with other primitive material on the smaller objects or within surface layers of the larger ones. They would be among the last things to melt and could survive if peak temperatures were low enough. In fact, the presence of CAIs during chondrule formation, close to the heat source, may be revealed by several facts: 1) some have an igneous texture which indicates they were melted, 2) dusty rims are present in some, which could have the same cause as the dusty rims of chondrules discussed below, 3) some chondrules have apparently absorbed some CAIs during their formation, as evidenced by an enhanced abundance of rare Earths, and 4) CAI-like relicts (if mostly only isotopically recognizable) may be present in some chondrules. \citep{K06, M03, Ma19}. The hybrid model described here does not appear to require any special storage history for CAIs or pre-solar grains. 

\subsection{Chondrite Lithification and Chondrule Age Spread}

Remarkably little has been written about the subject of chondrite lithification, especially when compared to the body of literature on chondrule formation. Lithification is the process of converting loose sediment into hard rock, which generally means reducing microporosity, increasing density and increasing tensile strength. It requires some degree of heat and pressure and may be facilitated by chemical processes. \citet{CB99} and \citet{CW02} have briefly elaborated on the basic problem -- there is not much heat or pressure available in the asteroid belt to accomplish lithification. It is generally assumed that the lithification of chondrites occurred on host bodies after the components (chondrules, matrix, CAIs, etc.) had accreted to them. It is also assumed that pressure and temperature spikes from impact shocks must have been, although there is little evidence for that in chondrites and the process has many problems \citep{BB16}. 

The hybrid model proposed here has two potential paths to chondrite lithification: it could have occurred on the solid portion of the merged planetesimal by a process mentioned in the paragraph above, or it could have occurred while the material was still in orbit by HIPping, as \citet{HG19} have proposed. In either case, metamorphism may have followed on the still-warm surface of the merged planetesimal as it continued to cool over millions of years. In the hybrid framework chondrites would be expected to contain mostly chondrules formed during a single collision event, so all the chondrules in a particular chondrite might be expected to be coeval. However, since a planetesimal could experience more than one chondrule-forming collision, it is conceivable that exceptions might exist.  While Pb-Pb ages do vary widely for chondrules within a single chondrite, the significance of those differences depends on the accuracy of the error estimates. Recent high precision age measurements based on Al-Mg chronology have not detected any significant age differences among chondrules within the same chondrite in a group of UOCs (unequilibrated ordinary chondrules) \citep{S22}, although \citet{T19} claim a spread of at least 1.1 Ma among chondrules in two CR meteorites.

\subsection{Relict Grains} 

Type I (Fe-O poor) chondrules often contain olivine grains with isotopic and minor element abundances that differ significantly from that of the majority of olivine grains in the same chondrule. In particular, the anomalous olivine grains are enriched in $^{16}$O and depleted in elements such as Ca, Al and Ti that are abundant in high temperature condensates such as CAIs and AOAs. They are known as relict grains \citep{R81, N81, J92, M18} because these attributes suggest that they are remnants of an original condensate from the hot gas that somehow avoided remelting during the formation of the chondrule. \citet{P21} have recently argued that precursor recycling and vapor-melt interactions were involved in the formation of relict grains found in both CC and OC chondrules. They describe a detailed chondrule formation scenario consistent with their high spatial resolution elemental and isotopic mapping of LL3 chondrules. It begins with an assemblage of primitive materials that are partially melted and evolve chemically and isotopically by interacting with their own vapor as they cool. More than one cycle of melting may be required. If this interpretation of relict grains is correct it is a strong argument that chondrules formed by melting primitive precursors, as proposed here. If the heating is primarily radiative, we might expect the temperature of material in the debris field to vary substantially on length scales of cms and timescales of minutes. This would be caused by the orbital motion of the debris, the highly variable attenuation resulting from self-shielding, and the possible dispersal of the heat source if there is a spray of hot material mixed with the primitive debris. High resolution simulations of low speed collisions are required to say anything more specific about what kind of heating/cooling curves would be expected but, again, the temperatures involved ($\ge$1600 K), the timescales ($\sim$30 minutes) and the expected variability on short length scales expected in a splash-flyby hybrid seem consistent with expectation based on observations of relict grains.

\subsection{Size and Size Sorting}

Chondrules have a remarkably narrow size distribution centered around 0.5 mm and small but significant differences between classes of different chemistry \citep{J14, F15}. A successful model of chondrule formation should account for that fact. If chondrules are born from sheets of expanding magma released during low speed collisions, as the splash model posits, their distinctive size could derive from a balance between surface tension of forming droplets and the ambient pressure of the gas, as \citet{A11} have suggested.  Presumably there could be a slight variation of conditions depending on chemical composition that could explain the observations. If chondrules formed by melting primitive solids one can also expect that drop formation will involve a balance between surface tension and ambient pressure, although the size and chemical composition dependence of radiative absorption efficiency could also affect the size distribution on its small end \citep{E95}. This is clearly an important issue that deserves more attention both theoretically and, perhaps, experimentally. No aerodynamic or other sorting of chondrules by size appears necessary in the hybrid model, since most or all of the chondrules in any chondrite would have formed together in the same location with the same chemistry under the same conditions and have a size distribution set by those conditions.  

\section{Summary}

We propose that chondrules formed when primitive debris from low-speed ($\le 1$ km s$^{-1}$) collisions of large (10-100 km radius) planetesimals, in the form of dust to m-scale boulders, was exposed to near-IR radiation from molten planetesimal interiors released by the collision.  The exposure could have been immediate, to magma ejected along with the primitive debris, and/or occurred later (and possibly repeatedly), as the solids orbited past exposed lava that remained on the surface of the merged object. This is a hybrid of the splash \citep{SS12, SS18} and flyby \citep{HG16, HG19} models. As \citet{A11}, \citet{W19} and \citet{A23} have discussed, low-speed collisions are expected to be common during the chondrule-forming epoch, 1-4 Ma after CAIs, as long as a proto-Jupiter was not yet affecting orbits in the asteroid belt. Typical collision velocities would be close to the escape speed of the larger planetesimals, around 450 km h$^{-1}$, comparable to what can be achieved by Formula 1 race cars. Such impacts are too slow, by a factor of about 25, to raise the temperature of material sufficiently to form chondrules by the ``jetting" mechanism \citep{JC18}, but they should generate substantial debris fields, similar to what was observed during the DART mission when the rubble-pile asteroid Dimorphos was impacted by a spacecraft \citep{JK23}. Testing this hypothesis will presumably require detailed numerical simulations of planetesimal collisions with high enough spatial resolution to follow the evolution of small-scale debris as it moves within an evolving thermal environment. As \citet{A23} notes, these collisions are complex and the parameter space large enough to accommodate a wide range of outcomes.

It remains an important task to firmly establish the setting or settings of chondrule formation, so that the vast store of information the chondrules harbor can be confidently decoded. For example, if the framework proposed here is correct then their magnetic properties may constrain a local field, not a global one. A collisional setting also suggests that, far from being the building blocks of planets, chondrules were never abundant in the asteroid belt, and today comprise $\le$ 0.1\% of its preserved primitive mass. In that case, we have just begun to tap the main repositories of primitive matter in the Solar System, through the Hayabusa2 and OSIRIS-REx missions, and should not be surprised if chondrules and chondrites continue to be rare or absent in returned samples. Asteroids and their spectrally-related meteorites clearly have similar chemical compositions, but based on bulk densities and inferred macroporosities (15\% or less), we expect the primary granular material of asteroids to be of lower density, lower tensile strength, less lithified and harboring many fewer chondrules than the chondrites in our museums.   

We thank Erik Asphaug for some helpful conversations and Ian Sanders and an anonymous referee for their insightful criticisms of the original version of this paper. We gratefully acknowledge the contributions of Ed Scott to the work presented here; it would not have been possible without him.

\end{document}